\documentclass[twocolumn]{openjournal}
\usepackage{natbib}
\usepackage{graphicx}
\usepackage{graphicx,amsmath,amssymb,amstext}
\usepackage{amsbsy,amsfonts,amsthm,color}
\usepackage{caption}
\usepackage{subcaption}
\usepackage{xcolor}
\usepackage{graphicx}
  \usepackage{lipsum}
\usepackage[colorlinks,linkcolor=blue,citecolor=blue,urlcolor=blue ]{hyperref}

\newcommand{\half}{\frac{1}{2}}

\newcommand{\cinv}{\ensuremath{ \sf{C}^{-1}}}

\newcommand{\bomu}{\ensuremath{\boldsymbol{\mu}}}

\newcommand{\boC}{{\ensuremath{\sf{C}}}}

\newcommand{\boSig}{\ensuremath{{\sf{\Sigma}}}}

\newcommand{\mathd}{\ensuremath{\mathrm{d}}}

\newcommand{\bop}{\ensuremath{\boldsymbol{p}}}
\newcommand{\boxi}{\ensuremath{\boldsymbol{\xi}}}

\newcommand{\calP}{\ensuremath{\mathcal{P}}}

\newcommand{\calL}{\ensuremath{\mathcal{L}}}

\newcommand{\calD}{\ensuremath{\mathcal{D}}}

\newcommand{\fatx}{\ensuremath{\boldsymbol{x}}}
\newcommand{\fatp}{\ensuremath{\boldsymbol{p}}}
\newcommand{\fatdelta}{\ensuremath{\boldsymbol{\Delta}}}

\newcommand{\sfpsi}{\ensuremath{{\sf{\Psi}}}}

\usepackage{soul}

\usepackage{xcolor}

\begin{document}

\title{Bayesian error propagation for neural-net based parameter inference}

\author{Daniela Grand\'on}
\email{daniela.grandon@ug.uchile.cl}
\affiliation{Grupo de Cosmolog\'ia y Astrof\'isica Te\'orica, Departamento de F\'isica, FCFM, Universidad de Chile, Blanco Encalada 2008, Santiago, Chile.\\Mathematical Institute, Leiden University, Snellius Gebouw, Niels Bohrweg 1, NL-2333 CA Leiden, The Netherlands.}

\author{Elena Sellentin}
\email{sellentin@strw.leidenuniv.nl}
\affiliation{Mathematical Institute, Leiden University, Snellius Gebouw, Niels Bohrweg 1, NL-2333 CA Leiden, The Netherlands.\\Leiden Observatory, Leiden University, Oort Gebouw, Niels Bohrweg 2, NL-2333 CA Leiden, The Netherlands.
}

\begin{abstract}
Neural nets have become popular to accelerate parameter inferences, especially for the upcoming generation of galaxy surveys in cosmology. As neural nets are approximative by nature, a recurrent question has been how to propagate the neural net's approximation error, in order to avoid biases in the parameter inference. We present a Bayesian solution to propagating a neural net's approximation error and thereby debiasing parameter inference. We exploit that a neural net reports its approximation errors during the validation phase. We capture the thus reported approximation errors via the highest-order summary statistics, allowing us to eliminate the neural net's bias during inference, and propagating its uncertainties. We demonstrate that our method is quickly implemented and successfully infers parameters even for strongly biased neural nets. In summary, our method provides the missing element to judge the accuracy of a posterior if it cannot be computed based on an infinitely accurately theory code.
\end{abstract}

\maketitle

\section{Introduction}
Cosmology currently finds itself in the fortunate state of collecting ever more, and in particular ever more precise data. The science-ready data sets of the upcoming surveys are forecasted to contain multiple million data points, yielding sub-percent precision on many cosmological parameters \citep{Euclid, LSST,SKA}. Likewise, the trend of collecting vaster, more precise data, is repeated in all other disciplines of astronomy that equally enjoy a progress in instrumentation.

If physical models and parameters are to be inferred from such data, then the increasing precision of data demands an equal increase in the precision of theoretical models. This need for improved theoretical modelling invalidates the formerly seen use of fast approximations, often resulting in numerical challenges that range from demanding to prohibitive. As artificial neural nets are -- once trained -- compellingly fast, astronomy lately sees a surge of neural-net based emulation of theoretical models and likelihoods.

\citet{ManriqueYus1} provide neural net predictions for half a million data points of a Euclid-like survey, computing the posterior of seven cosmological parameters when also emulating the theoretical predictions for a Planck-like \citep{Planck2018} observation of the cosmic microwave background. \citet{cosmopower} provide a neural net emulator of the cosmic matter powerspectrum, which facilitates the incorporation of nuisance parameters. In \cite{serenaemulator}, the authors provide neural net based molecular line modelling for inferences in astrochemistry. Neural net based inference has also been demonstrated on simulated LSST-year1 data in \citet{NNLSST}.

Common to all of these examples is that the respective authors demonstrate the validity of their neural-net based inference by comparing to the posterior computed with a classical non-approximative code. In an actual research case, this comparison will often be unfeasible: if the neural-net assisted posterior is the only one that can be computed in reasonable time, then no other posterior is available for comparison. It would then be uncertain whether the neural net had the required accuracy to enable an unbiased inference.

Accordingly, a thus far unsolved problem in neural-net assisted inference was how to avoid that the neural net's approximation error will bias the parameter inference. As any neural net is by setup an approximator, it is vital to have a clean solution for propagating its error. 

This paper hence presents an adapted posterior function, which is the Bayesian solution to the wish of wanting to eliminate a neural net's bias and propagating the uncertainty of its prediction. Crucially, our solution is expected to work for a wide range of successfully and rather unsuccessfully trained neural nets. As the neural net training improves, our adapted posterior converges to the posterior yielded by traditional non-neural net based inferences.

Our paper is structured as follows. The Bayesian solution to propagating biases and approximation errors in a neural-net assisted inference is presented in Sect.~\ref{Posterior}. This papers' main result are Eq.~(\ref{general1}) and Eq.~(\ref{general2}), which give the adapted posterior function that replaces the posterior yielded from a non-approximate code. Eq.~(\ref{general1}) is the special case of Eq.~(\ref{general}) if the data follow a Gaussian sampling distribution and the neural net's errors can be approximated as Gaussian too. Sect.~\ref{Inference} compares this adapted posterior to the `naive' posterior gained from wrongly assuming the neural net has an infinite precision. We demonstrate two cases of varying neural net training success. Our results are briefly discussed in Sect.~\ref{Discussion}.

\section{Bayesian posterior for neural-net assisted inference}
\label{Posterior}

We assume a neural net that has been trained to output predictions $\bomu_{\rm NN}(\bop)$, and has undergone validation by testing its predictions on an independent validation set that contains $V$ samples of parameters $\bop$.

The neural net's approximation error can then be evaluated at each sample of this validation set, and we define the discrepancy between the neural net prediction and approximation-free mean of the data at $\bop$ (for example, the observable output of a Boltzmann solver in cosmology) to be $\fatdelta (\bop) = \bomu(\bop) - \bomu_{\rm NN} (\bop)$. However, in the space between the validation points, the approximation error will remain unknown. Hence, as $\fatdelta(\bop)$ is unknown for all but a few points in parameter space, we promote it to a random, since uncertain, variable in the Bayesian sense. We shall therefore treat $\fatdelta(\bop)$ as random for the same reason that the unknown covariance matrix of \citet{SH} is treated as random.  

For any random variable, a Bayesian solution requires the probability distribution that this variable follows. We denote the distribution as $\fatdelta(\bop) \sim \calD(\fatdelta(\bop)| \cdot)$, where the placeholder $\cdot$ denotes variables that specify the distribution further, for example, a mean and a variance if $\calD$ is a Gaussian.

In the worst case of neural net training, the neural net will have approximation errors that strongly vary with the values of $\bop$. For example there might be parameter ranges for which the neural net succeeds in approximating the parametric explanation of the data, and there might be areas where it deviates strongly from the training set. In this case, we would have $\calD(\fatdelta(\bop)|\bop, \cdot)$, i.e. the distribution of neural net approximation errors will explicitly depend on the parameters themselves. A Bayesian solution to propagating the neural net's errors would then still be possible, as long as $\calD$ can be accurately estimated. This could e.g. be achieved by histograming the approximation errors over different patches of the parameter range. Another solution to a spatially-varying distribution $\calD$ would be to model it with a Gaussian process, see e.g.~\citet{HarryGauss}, who however find that a Gaussian process is an inaccurate error model for a deterministic function. Accordingly, they find that the posterior using a Gaussian process model differs more from the true posterior than if the error were ignored all together. 

Adopting a Gaussian process would also allow us to treat the approximation error as known at all validation points. However, using a Gaussian process for the sake of modelling spatially varying approximation errors necessitates that hyperparameters must be set. To avoid such hyper-parameters, we hence prefer to treat the approximation error as random also at the validation points. This is a conservative choice as it treats the approximation error at the validation points with the same uncertainty as in between the validation points.

We hence adopt the approximation that for all values of $\bop$, the neural net's errors follow the same distribution, namely $\fatdelta \sim \calD(\fatdelta|\cdot)$, where the placeholder $\cdot$ now excludes the parameters $\bop$. This working hypothesis could be dropped at the expense of a more involved derivation, but in numerical experiments of reasonably trained nets we found it to hold with excellent precision.

\subsection{Marginalizing the neural net error and bias}
We assume the neural net provides parametric predictions for $n$ data points, and we denote the scientific data by $\fatx$\footnote{The scientific data are not the training set.}. The likelihood of the data $\calL(\fatx|\bop, \fatdelta)$ is dependent on the neural net's approximation error $\fatdelta$: if the neural net's prediction is erroneous, other values of the parameters $\bop$ might fit the data better in order to compensate for the neural net's error. Such a bias in the inferred parameters is the key worry we here aim to eliminate by accounting for the neural net's error.

We also assume that the prior probability of the parameters $\pi(\bop)$ is independent of the prior probability of approximation errors $\pi(\fatdelta)$. This will hold if the neural net was trained according to best practice, in particular if it was provided with `sufficiently' many training points in each subvolume of the prior range. In brief, the training has to guarantee that $\pi(\bop,\fatdelta)$ factorizes. Counterexamples can be engineered. Namely $\pi(\bop,\fatdelta)$ will not factorize when
the neural was trained on highly unrepresentative samples of $\bop$, for example having no training samples in certain subvolumes, or if $\fatdelta$ particularly deviates from average in certain region of parameter space.

Propagating the neural net's prediction error then implies that we must marginalize $\fatdelta$.  The posterior of parameters inferred from an imperfect neural net prediction is hence the compound of the posterior of vanishing neural net error with the distribution of the error
\begin{equation}
\begin{aligned}
    \calP(\bop|\fatx) & = \int \calP(\bop,\fatdelta|\fatx) \mathd^n\Delta\\
   & = \int_{-\infty}^{+\infty} \calP(\bop|\fatdelta,\fatx)\calD(\fatdelta|\cdot)\mathd^n\Delta\\
   & \propto \int_{-\infty}^{+\infty} \calL(\fatx|\bop, \fatdelta) \pi(\bop) \calD(\fatdelta|\cdot)\mathd^n\Delta,
    \end{aligned}
    \label{general}
\end{equation}
where the proportionality sign arises by not having written out the normalization with the Bayesian evidence. 

Eq.~(\ref{general}) describes that parameter inference can be safeguarded against a neural net's approximation error by specifying the data's likelihood as a function of physical parameters and the net's approximation error, setting a prior on the parameters, and compounding with the distribution of the net's approximation error. The posterior of physical parameters is hence the marginal over the neural net's approximation error.

We shall now specialize Eq.~(\ref{general}) to the most common case of Gaussian data.

\subsection{Specialization to Gaussian data}
\label{posterior_gaussian}

In this section we provide a closed-form expression for the posterior of marginalized neural net error for the case of Gaussianly distributed data.

In line with this assumption, the likelihood now reads
\begin{equation}
    \calL(\fatx | \fatp) \propto \exp\left(-\half [\fatx - \bomu(\fatp)]^T \cinv [\fatx - \bomu(\fatp)] \right),
\end{equation}
where $\bomu$ is the error-free mean of the data, evaluated as a function of the parameters $\bop$, and $\cinv$ is the inverse covariance matrix of the data. 

Defining $\boxi(\bop) = \fatx - \bomu(\bop)$, the likelihood with neural net error is
\begin{equation}
    \calL(\fatx | \fatp, \fatdelta) \propto \exp\left(-\half [\boxi(\bop) - \fatdelta]^T \cinv [\boxi(\bop) - \fatdelta] \right).
    \label{GaussL}
\end{equation}

We define the mean of the neural net approximation error as
\begin{equation}
    \bar{\fatdelta} = \frac{1}{V} \sum_{v = 1}^V \fatdelta_v
    \label{fatdelta}
\end{equation}
were $v$ runs over the samples of the validation set. Accordingly, the covariance matrix of the neural net's prediction error is
\begin{equation}
    \boSig = \frac{1}{V-1} \sum_{v =1}^V (\fatdelta_v - \bar{\fatdelta})(\fatdelta_v - \bar{\fatdelta})^T.
    \label{fatsigma}
\end{equation}
For an excellently trained neural net, the mean $\bar{\fatdelta}$ will vanish, meaning the neural net provides an unbiased prediction of the theory function over the entire parameter range. On the other hand, the covariance matrix $\boSig$ will contain large variances and covariances for a neural net whose training did not yield very accurate predictions of the target function. The more training samples were seen, and the smaller the loss during training, the smaller the elements of $\boSig$ will become. 
The Gaussian distribution of $\fatdelta$ then is
\begin{equation}
    \calD(\fatdelta) \propto \exp\left( -\half [\fatdelta-\bar{\fatdelta}]^T \boSig^{-1} [\fatdelta -\bar{\fatdelta}] \right).
    \label{GaussD}
\end{equation}
In the limit $\boSig \to 0$, this distribution tends to Dirac's delta distribution, correctly reflecting that no errors are folded into the posterior of parameters.

As the compound of a Gaussian with another Gaussian yields a third Gaussian, inserting Eq.~(\ref{GaussL}) and Eq.~(\ref{GaussD}) into Eq.~(\ref{general}) yields
\begin{equation}
    \calP(\bop|\fatx) \propto \exp\left(-\half [\boxi(\bop) - \bar{\fatdelta}]^T \sfpsi [\boxi(\bop) - \bar{\fatdelta}] \right) \pi(\bop),
\label{general1}
\end{equation}
where
\begin{equation}
    \sfpsi = ( \boC + \boSig)^{-1},
    \label{general2}
\end{equation}
and $\bar{\fatdelta}$ is given by Eq.~(\ref{fatdelta}). We expect Eq.~(\ref{general1}) to be the result of this paper that is at minimum effort most widely applicable: For a neural net that produces approximately Gaussian distributed inaccuracies in its predictions, Eq.~(\ref{general1}) propagates the net's uncertainties, and eliminates the net's bias due to the subtraction of $\bar{\fatdelta}$. Eq.~(\ref{general1}) hence eliminates neural net biases in the inference of cosmological parameters.

\section{Demonstration of parameter inference}
\label{Inference}
In this section we demonstrate the inference of parameters when the theory prediction for the data has been computed with a neural net. We will highlight two cases of how (un)successful neural net training deforms the posterior of inferred parameters.

For our demonstration, we simulated a cosmological analysis of the matter power spectrum $P(k)$ as a function of wavenumber $k$. Our data range spans $k \in [0.0033, 0.4]$ in units of h/Mpc. In order to train a neural net, we generate 10000 power spectra with CAMB \cite{Lewis:2002ah} by randomly drawing from the prior ranges given in Table~\ref{table1} for the parameters $H_0, \Omega_{\rm cdm}, \Omega_{\rm b}$, which are the Hubble constant, the density parameter of cold dark matter, and the density parameter of baryons, respectively. We split the CAMB predictions into a batch of 70\% for training, 20\% for the validation set and 10\% for test. These batches are used to find the best architecture for the neural nets, where the test set allows to validate
the performance of the neural network on an unseen set of points.
We also segmented the power spectra into wavenumber ranges of $k \in [0.0033, 0.044]$ and $k \in [0.044,0.40]$ and trained a neural net for each $k$-range individually. The segmentation into $k$-ranges is not necessary, but advantageous because it reduces the layer sizes and thereby the trainable parameters.

\setlength{\tabcolsep}{10pt}
\renewcommand{\arraystretch}{1.5}

\begin{table}[t]
\centering
\begin{tabular}{||c c c||} 
 \hline
  & Prior ranges & Fiducial cosmology \\ [0.5ex] 
 \hline
 $\Omega_{\rm cdm}$ &  [0.22,0.4] & 0.269\\
 
 $\Omega_{\rm b}$ & [0.02,0.06] & 0.0494\\

 $H_0$ & [58,82] & 66.8\\
 \hline
\end{tabular}
\caption{Prior ranges and input cosmology for $\Omega_{\rm cdm}$, $\Omega_{\rm b}$ and $H_0$.}
\label{table1}
\end{table}

For the training we engineered two cases. One is a successfully/accurately trained neural net whose error on the test set is at the percent level on average over k. The second case is an unsuccessfully/inaccurately trained neural net where the prediction error is up to 40\%. Such big an error would clearly bias the posterior of cosmological parameters unless it is corrected for. Further details about the performance of our neural nets are presented in Appendix~\ref{appendixa}. 

To proceed with the Bayesian error propagation, we generate a fourth set $\mathbf{V}$ including new 10000 power spectra at random $\bop$ from the prior range. We found it advantageous if the fourth set includes more points than the test set.
For each point $v$ in $\mathbf{V}$ it is possible to obtain the neural net approximation error $\fatdelta_v(\bop) = \bomu(\bop) - \bomu_{\rm NN} (\bop)$ that accounts for the bias in the prediction, and $v$ runs from 1 to 10000. 

\begin{figure}
     \centering
     \begin{subfigure}[b]{1.0\linewidth} \hspace*{-1.2cm} 
         \includegraphics[width=1.2\textwidth]{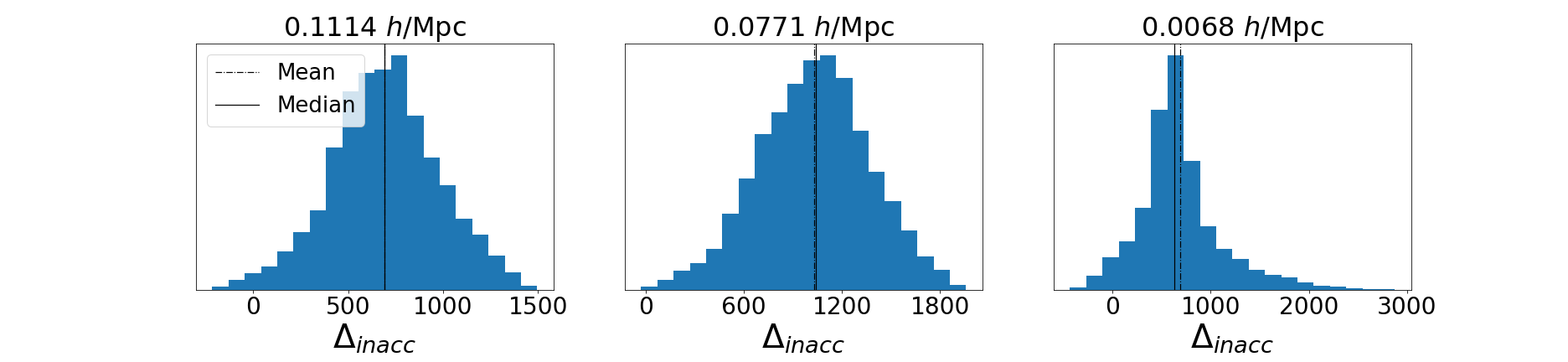}
         \caption{Histograms obtained from the inaccurate model}
         \label{fig: sg}
     \end{subfigure}
     \begin{subfigure}[b]{1.0\linewidth}
     \hspace*{-1.2cm} 
         \includegraphics[width=1.2\textwidth]{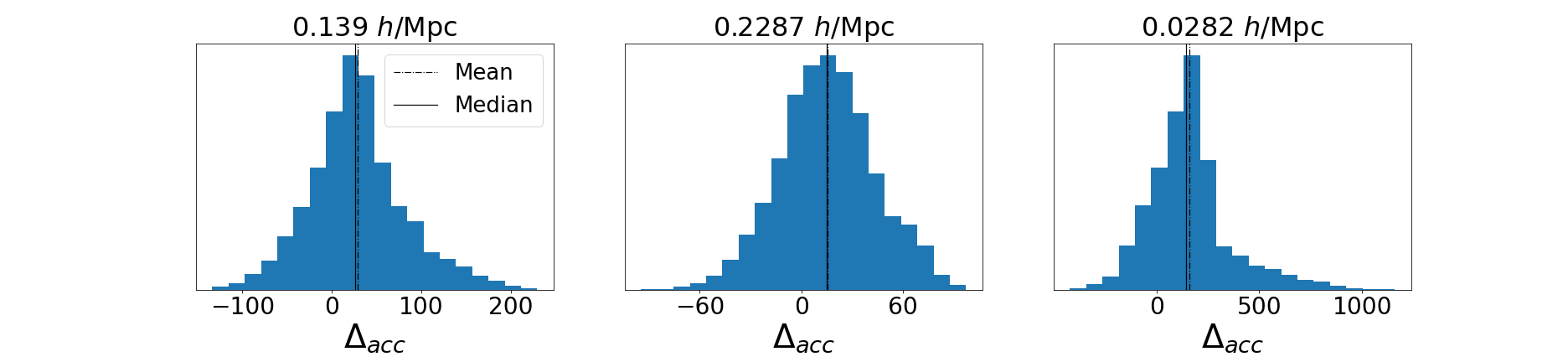}
         \caption{Histograms obtained from the accurate model}
         \label{fig: sw}
     \end{subfigure}

     \caption{Histograms of $\fatdelta$ at specific wave number $k$ obtained for all the points in fourth set. From left to right, the two first histograms represent the general tendency of our histograms, where gaussianity condition is fullfilled. The third column shows histograms where the distributions slightly deviate from gaussianity.}
        \label{gaussians}
\end{figure}

In this paper, as stated in Section \ref{posterior_gaussian}, we assume $\fatdelta$ is a random variable that follows a Gaussian distribution, so it is crucial to check this assumption by computing histograms for different $k$. Fig.\ref{gaussians} shows histograms for the two models at different $k$. As a very basic test, the mean and the median are displayed as we expect both match for a normally distributed sample. We also computed skewness and curtosis and ensured that no bimodal structure occurs in the histograms. By computing these higher-order moments and applying Kolmogorov-Smirnov test and p-value to all histograms of k, we found that our assumption of Gaussianity is indeed supported for both models. On the other hand, we check that $\calD(\fatdelta)$ does not depend on parameters by doing a sub-volumes analysis, where we split parameter in space in different patches and compute the mean of the histograms. All details can be found in Appendix~\ref{appendixb}.

For both upper and lower panels of Fig.~(\ref{gaussians}), the first two figures from left to right show cases where gaussianity assumption is completely fullfilled. The third column shows handpicked examples where the gaussian approximation is stretched, but is a very rare case and no more extreme cases were observed.

\begin{figure}
\centering 
\includegraphics[width=0.5\textwidth]{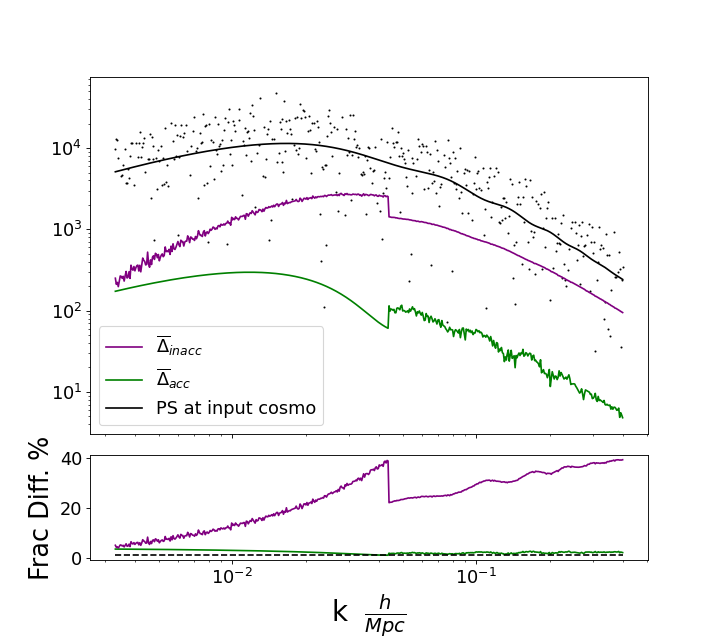}
\caption{The prediction errors for the two neural nets are displayed and compared to the matter power spectra at the input cosmology. The prediction error for the accurate model $\bar{\fatdelta}_{\rm acc}$ is shown in green and the prediction error $\bar{\fatdelta}_{\rm inacc}$ for the inaccurate model in purple. The scatter plot represent the synthetic matter power spectra used for the inference.}
\label{deltas}
\end{figure} 

\begin{figure}
\hspace*{-1.1cm} 
\includegraphics[width=0.62\textwidth]{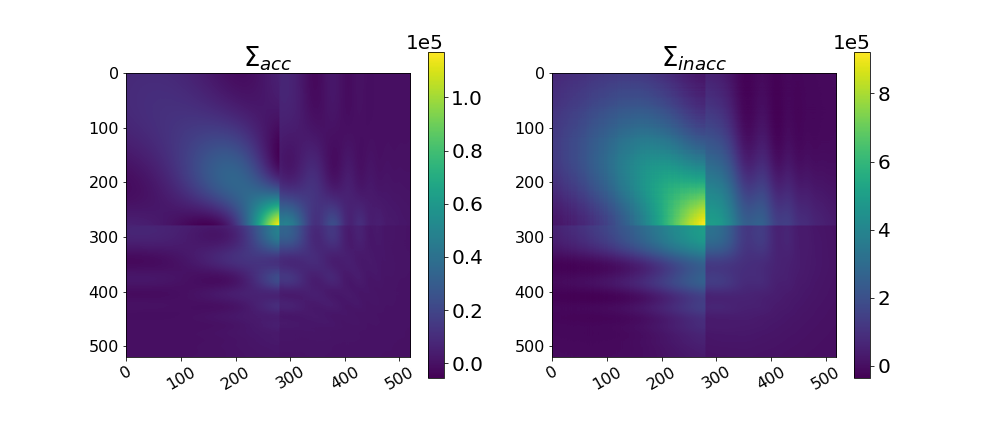}
\caption{Left: the covariance matrix $\Sigma_{\rm acc}$. Right: $\Sigma_{\rm inacc}$. We see that the badly trained neural net causes not only a higher bias as seen in Fig.~(\ref{deltas}) but additionally also more correlated errors in its prediction.}
\label{sigmas}
\end{figure} 

As presented in Eq.~(\ref{general1}), our adapted posterior accounts for the bias and the appoximation error by defining $\bar{\fatdelta}$ and $\boSig$. From $\fatdelta_v(\bop)$, we apply Eq.~(\ref{fatdelta}) and  Eq.~(\ref{fatsigma}) to obtain {$\bar{\fatdelta}_{\rm acc}$,$\boSig_{\rm acc}$} and {$\bar{\fatdelta}_{\rm inacc}$,$\boSig_{\rm inacc}$} for the accurate and inaccurate models, respectively. In this demonstration, the quantities $\fatdelta$ and $\boSig$ (for both accurate and innacurate models) do not vary either using 8000, 9000 or 10000 points of V. Also, the sub-volumes analysis in Appendix~\ref{appendixb} shows that the approximation error is constant throughout parameter space. Hence, we can safely state that $\bar{\fatdelta}_{\rm acc}$ and $\bar{\fatdelta}_{\rm inacc}$ represent the approximation error of the neural net's for this particular case.
The prediction errors $\bar{\fatdelta}_{\rm acc}$ and $\bar{\fatdelta}_{\rm inacc}$ are displayed in Fig. \ref{deltas}, and compared to the power spectra at the input cosmology. The Lower panel shows the fractional difference that goes up to 5\% compared to the input power spectra for the accurate neural net, and 40\% for the inaccurate model. Hence, we expect the bias of the best fit of parameters is negligible when comparing the resulting posteriors with and without error propagation for the accurate model. In Fig. \ref{sigmas} we see the variance and covariances for the unsuccessfully trained model are larger, hence impacting strongly the covariance matrix of the adapted posterior. 

Once all the components of Eq.~(\ref{general1}) are derived, we constrain cosmological parameters for three scenarios: when the theory prediction is computed by CAMB, and when the prediction is done with the accurate neural net model, and inaccurate model. Moreover, for neural-net based inference we consider the cases when the posterior does and does not propagate the neural net's error, where the former is the adapted posterior proposed in Eq.~(\ref{general1}) and the latter a naive plug-in of the neural net into the standard Gaussian likelihood.

\begin{figure}
\centering 
\includegraphics[width=0.4\textwidth]{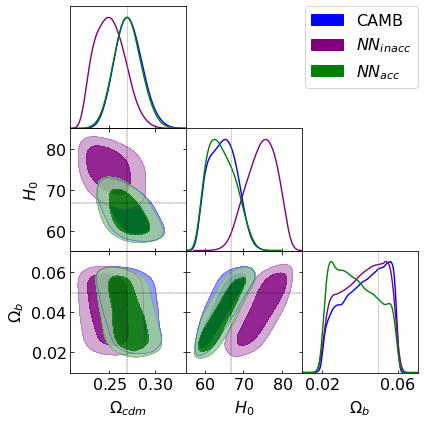}
\caption{Marginal posterior contours obtained for neural-net based inference and normal Boltzmann-code based inference for synthetic matter power spectra data. The contours contain 68\% and 90\% posterior volume.}
\label{wnn}
\end{figure}   
 
We generated synthetic matter power spectra by Gaussian random draw from a mean, depending on input parameters $H_0 = 66.8$ km s$^{-1}$ Mpc$^{-1}$, $\Omega_{\rm cdm} = 0.269, \Omega_{\rm b} = 0.0494$, and a covariance matrix $\boC$ as the square of the fiducial power spectrum. The mean is displayed as a black solid line in Fig.~\ref{deltas}, where the synthetic data scatters around.

In all figures, purple contours stand for the posteriors obtained from the inaccurate neural net, and green posteriors for the accurate net. The input parameters are indicated in all corner plots by dashed black lines. In Fig. \ref{wnn}, we depict the `naive' posteriors, yielded by simply replacing the (taken as infinitely accurate) CAMB mean $\bomu(\bop)$ with the neural net's prediction, neither propagating the neural net's bias nor its approximation error around the true mean. Purple contour show a noticeable non-zero bias in the best fit, while the green contour mostly match the CAMB blue contour. Hence, Fig.~\ref{wnn} demonstrates that parameter inference will incur biases if the neural net's approximation error is not propagated.

\begin{figure}
\centering 
\includegraphics[width=0.4\textwidth]{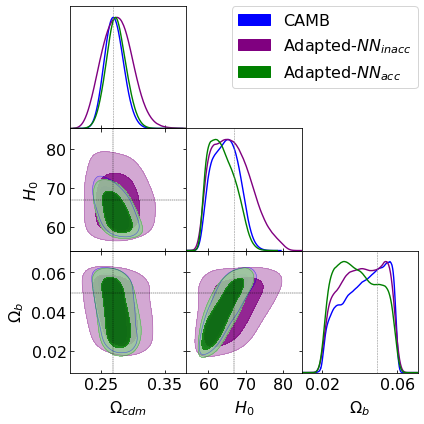}
\caption{Marginal posterior contours obtained for neural-net based inference implementing the proposed adapted posterior. CAMB results are also displayed for comparison. The contours contain 68\% and 90\% posterior volume.}
\label{gnn}
\end{figure}

Accordingly, Fig.~\ref{gnn} demonstrates the main results of this work, namely the adapted-posterior that propagates the neural nets errors into the final parameter inference. For both neural nets, the posterior of parameters is widened up due to the neural nets approximation error. This is a consequence of the covariance matrix of the neural net approximation error shown in Fig. \ref{sigmas}. As this covariance matrix adds to the original covariance matrix of the data, the uncertainty on the parameter constraints increases. This is much more prominent for the inaccurately trained neural net. Also seen is that the subtraction of the neural net's bias leads to a debiasing of the posterior: The fiducial cosmology is included in the posterior after debiasing, but is excluded beforehand. This evidences that our Eq.~(\ref{general1}) is a simple but powerful method for safeguarding parameter inference against the approximative nature of neural nets.

\section{Discussion}
\label{Discussion}

In this work we derived an adapted posterior that propagates the error of the theory predictions of  neural-net based inference. The adapted posterior results in Eq. \ref{general1} and  \ref{general2}, and implies a modified precision matrix and a bias correction when the error in the neural net prediction is non-negligible. We investigate the impact of the neural net prediction error on a cosmological analysis for constraining parameters $\Omega_{\rm cdm}$, $\Omega_{\rm b}$ and $H_0$. 
The method is easy to implement when the prediction errors follow a gaussian distribution, which must be checked. It is worth noting that for this approximation gaussianity holds even if a few examples present mild longer tails. However, non-negligible higher moments can emerge in cases where the distribution is bimodal or where heavy tails are present. In those cases it is crucial to either extend the analysis and propose a new analytic form for $\calD(\fatdelta)$, or to improve the neural net training, in order to achieve Gaussian approximation errors. In this analysis, we found that gaussianity holds in most of the trained models, making the method simple to apply.

In Section \ref{Inference} we show that error propagation can restore the input cosmology in the best-fit even if the neural net is inaccurate. 

As many works in the literature had shown the advantages of accelerated neural-net based inference in the context of future Stage-IV cosmological surveys, we consider it critical to also include error propagation for neural nets inference. In the era of precision cosmology, where many efforts are still being done on the control of astrophysical or instrumentation systematic effects, it is also necessary to consider and characterise the biases that our modern statistical methods propagates into the inference.\\

\vspace{0.5cm}

\begin{figure}
\hspace*{-1.1cm}
\centering 
\includegraphics[width=0.62\textwidth]{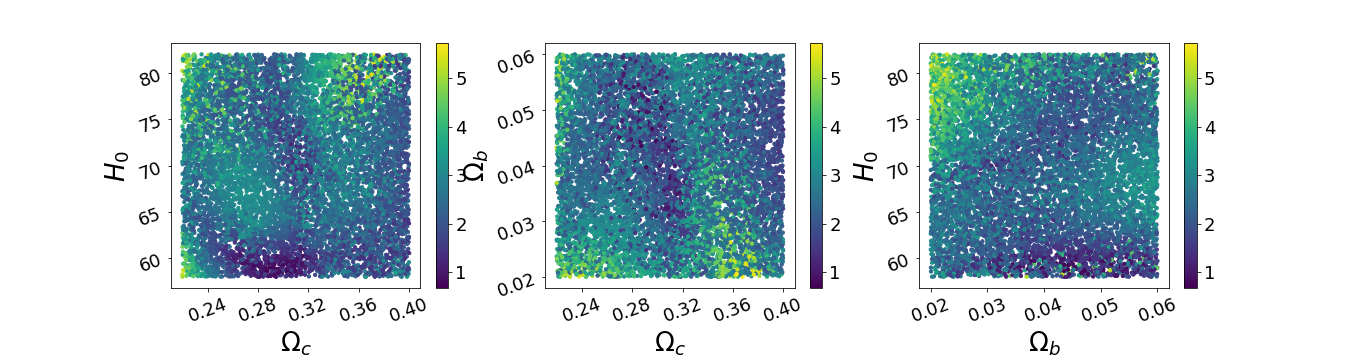}
\caption{Scatter plot of parameter space where the colorbar stands for the average (over all wave numbers) percentage error in the prediction made by the accurate neural net.}
\label{scat_gnn}
\end{figure}

\begin{figure}
\hspace*{-1.cm}
\centering 
\includegraphics[width=0.61\textwidth]{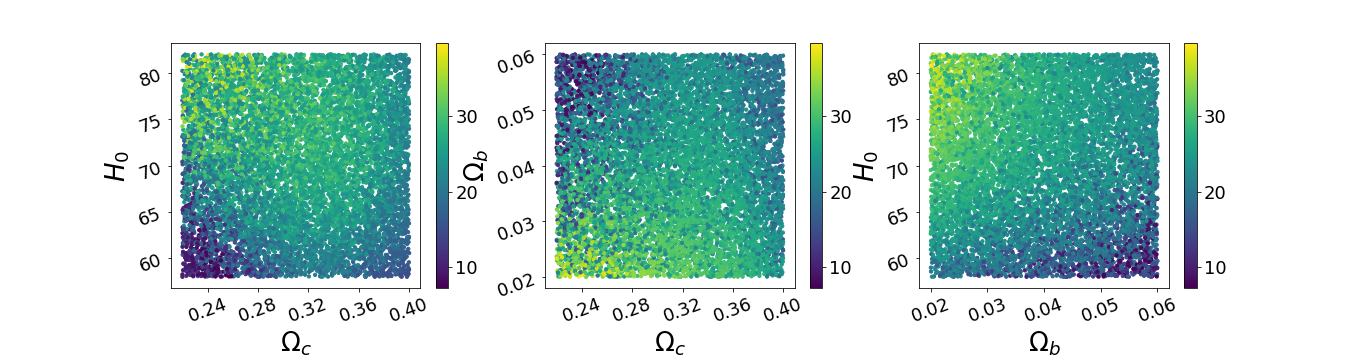}
\caption{Same as Fig.~(\ref{scat_gnn}) but for the innacurate neural net.}
\label{scat_wnn}
\end{figure}

\section*{Acknowledgements}

We thank the support staff of Leiden’s ALICE High Performance Computing
infrastructure. DG acknowledges financial support 
by project ANIDPFCHA/Doctorado Nacional/2019-2119188 and thanks the Mathematical Institute of 
Leiden University for the hospitality during this research.

\appendix

\section{Appendix A}
\label{appendixa}

Our neural nets are validated over a 10000 points fourth set, which is also used to compute $\bar{\fatdelta}$ and $\boSig$ that define our new posterior. Fig.~\ref{scat_gnn} and Fig.~\ref{scat_wnn} show the performance of the models on the fourth set. Each point in parameter space displays the average error over all wavenumbers.  

As can be seen, the accurate net mainly produces 3\% average error with a few points going up to 6\%. On the other hand, the bulk of the points for the inaccurate model reach 30\% errors.

\section{Appendix B}
\label{appendixb}

\begin{figure}
\hspace*{-1.1cm}
\centering 
\includegraphics[width=0.62\textwidth]{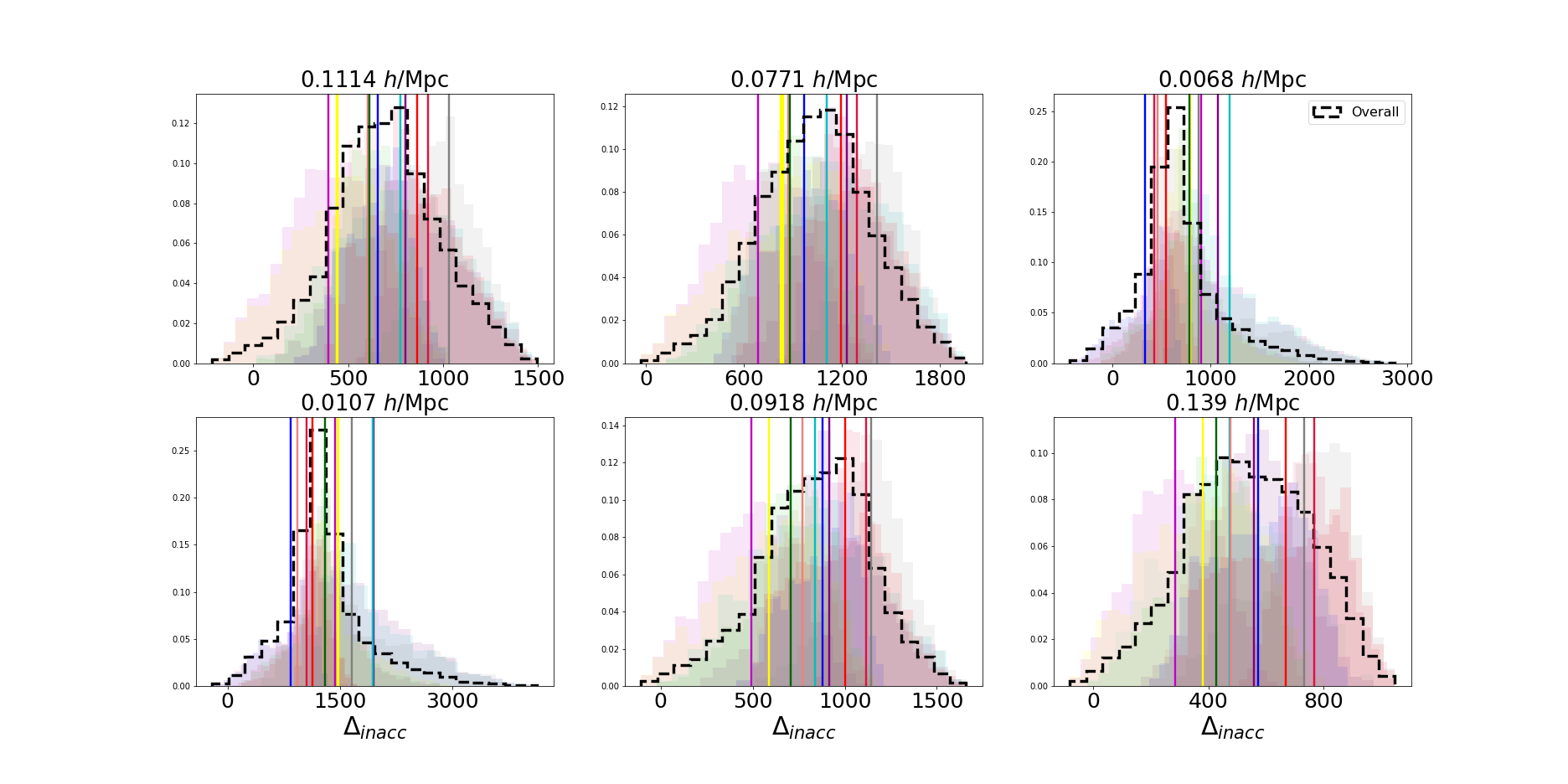}
\caption{Sub-volume histograms for different wavenumber $k$, where $\fatdelta$ is obtained with the innacurate neural net. The overall distributon over all parameter space is depicted in a dashed-black contour. Shaded colors in the background represent the obtained histograms for different chosen sub-volumes, where the means are shown in vertical coloured lines. We conclude that the neural net approximation error does not directly depend on the sub-volume and hence not directly on the cosmological parameters themselves.}
\label{subv_wnn}
\end{figure}

\begin{figure}
\hspace*{-1.1cm}
\centering 
\includegraphics[width=0.62\textwidth]{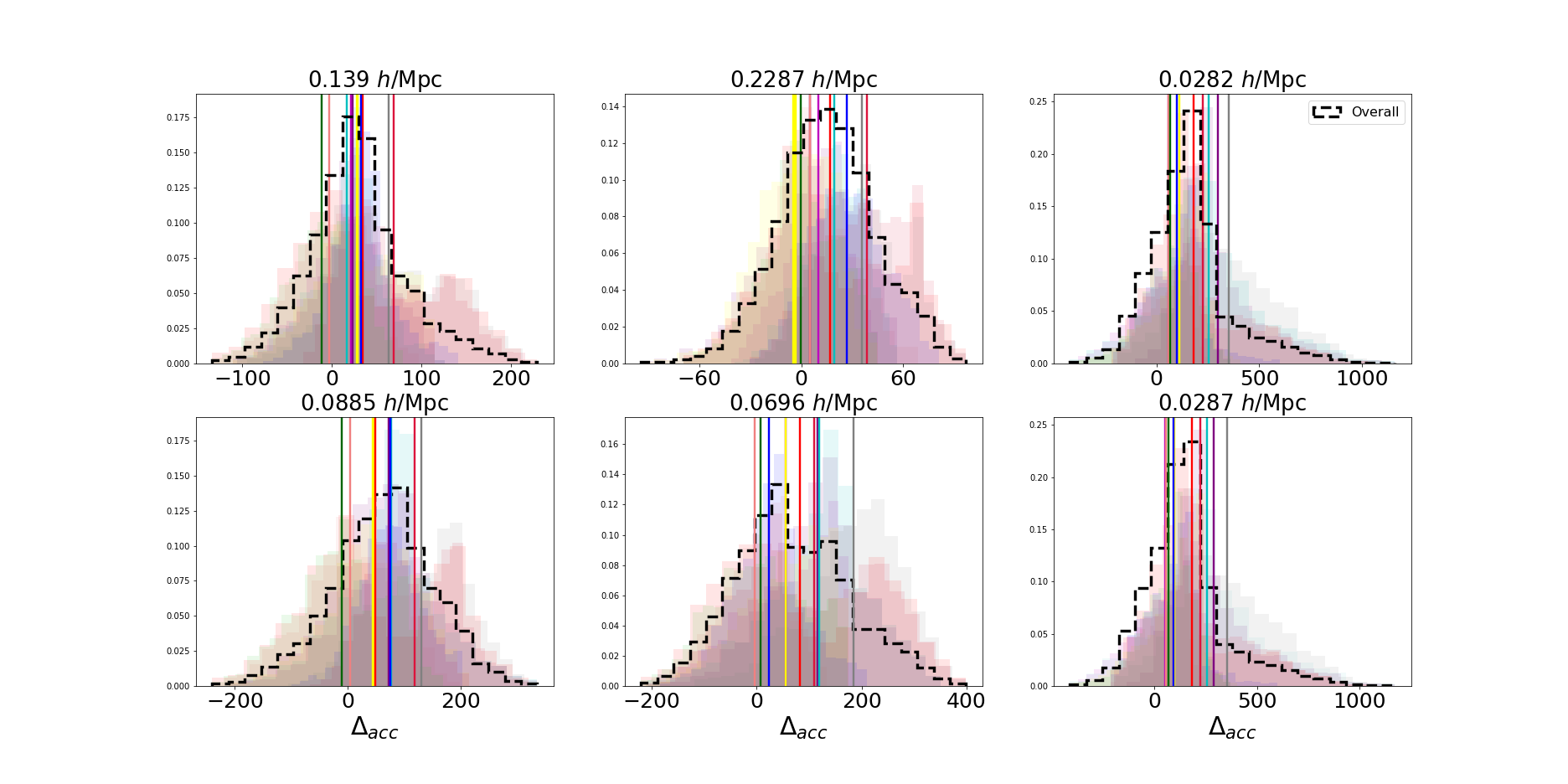}
\caption{As Fig.~(\ref{subv_wnn}) but for the accurate neural net, with the same conclusion.}
\label{subv_gnn}
\end{figure}

In order to further validate our assumption that $\calD(\fatdelta)$ is Gaussian and constant throughout parameter space, we do a sub-volume analysis of this distribution. By sub-volume we imply small regions of the parameter space, where we study if Gaussianity also holds and if the mean $\fatdelta$ in these sub-volumes falls well within the bulk in the general $\calD(\fatdelta)$ distribution. In Fig.~\ref{subv_wnn} and Fig.~\ref{subv_gnn} we see that most of the sub-volume means fall into the 2$\sigma$ region and this holds for both the accurate and the inaccurate neural net. Hence, $\calD(\fatdelta)$ does not depend on cosmological parameters which fulfills our assumption.

We further tested in each subvolume the assumption of Gaussianity by computing the skewness and curtosis and found that Gaussianity is verified with sufficient accuracy. In particular, there were no bimodal distributions for $\fatdelta$, which would most quickly invalidate the approximation of $\fatdelta$ being Gaussianly distributed.

\bibliographystyle{mnras}
\bibliography{GrandonSellentin.bib}

\end{document}